%% file: paper.tex
\documentclass{llncs}

\usepackage{subcaption}
\usepackage{adjustbox}
\usepackage{fancyvrb}
\fvset{frame=single,framesep=1mm,fontfamily=courier,fontsize=\scriptsize,numbers=none,framerule=.2mm,numbersep=1mm,commandchars=\\\{\}}
\usepackage{xspace}
\usepackage{times}
\usepackage{booktabs}
\usepackage[scaled=0.92]{helvet}
\usepackage[T1]{fontenc}
\usepackage{amssymb}
\usepackage{textcomp}
\usepackage{graphicx}
\usepackage[breaklinks=true]{hyperref}
\usepackage{float}
\usepackage{etoolbox}
\usepackage{url}
\usepackage{mathptmx}
\usepackage[noadjust]{cite}
\usepackage{mathtools}
\usepackage{stackengine}
\setstackEOL{,}
\DeclarePairedDelimiterX{\abs}[1]\lvert\rvert{\ifblank{#1}{\,\cdot\,}{#1}}
\title{Placement Laundering and the Complexities of Attribution in Online Advertising}
\titlerunning{Placement Laundering}
\author{Jeffery Kline\inst{1} \and Aaron Cahn\inst{1} \and Paul Barford\inst{1,2}}
\institute{comScore Labs \and University of Wisconsin--Madison}
\begin{document}
\maketitle
\begin{abstract}
  \input{abstract}
\end{abstract}

\providecommand{\myvspace}{\vspace{-1.5ex}}
\providecommand{\vsection}[1]{\myvspace \section{#1} \myvspace}
\providecommand{\vsubsection}[1]{\myvspace \subsection{#1} \myvspace}

\setlength\emergencystretch{\hsize}

\vsection{Introduction} \label{sec:intro} \input{intro}
\vsection{Ad Ecosystem Overview} \label{sec:background} \input{background}
\vsection{Methods of Attribution and Placement Laundering} \label{sec:stdmeth} \input{stdmeth}
\vsection{Case Studies} \label{sec:casestudies} \input{casestudies}

\vsection{Detection and Measurement} \label{sec:detectmeasure} \input{detectmeasure}

\vsection{Discussion} \label{sec:discuss} \input{discuss}
\vsection{Related Work} \label{sec:relwork} \input{relwork}

\vsection{Summary and Conclusions}\label{sec:sumry} \input{sumry}
\clearpage
\bibliographystyle{splncs}
\bibliography{paper}
\vsection{Appendix: Bimage Case Study}\label{sec:appendix}\input{appendix}
\end{document}

%% file: abstract.tex
A basic assumption in online advertising is that it is possible to attribute a 
view of a particular ad creative ({\em i.e.,} an impression) to a particular 
web page.  In practice, however, the seemingly simple task of ad attribution is 
challenging due to the scale, complexity and diversity of ad delivery systems.  
In this paper, we describe a new form of fraud that we call {\em placement 
laundering} which exploits vulnerabilities in attribution mechanisms. 
Placement laundering allows malicious actors to inflate revenue by making ad calls that appear to originate from high quality publishers.
We describe the basic aspects of placement laundering and give details of two 
instances found in the wild.  One of the instances that we describe abuses
the intended functionality of the widely-deployed SafeFrame environment.
We describe a placement laundering detection method that is capable of identifying a general class of laundering schemes, and provide results on tests with that method.

%% file: intro.tex
Advertising is one of the primary revenue sources in the Internet today.   
Beyond the fact that people are spending increasing amounts of time online, 
advertising on the Internet is distinguished by virtue of the ability to target 
users and to quantify the impact of ad spend through key metrics such as 
click-through and conversion.  Recent reports by the Interactive Advertising 
Bureau (IAB) place online ad spend at \$40.1B for the first half of 2017 in the 
US and growing at an annual rate of 22.6\%~\cite{IAB}.  These revenues come 
from billions of ads delivered to millions of users every day.

The process of delivering ads at Internet scale on a daily basis has many challenges.  When a user accesses a web page or uses an app that includes space set aside for ads (this is the ad's {\em placement}), vast distributed infrastructures are invoked, which facilitate delivery in a timely fashion (typically under 300ms).  The delivery of an ad to a user who views the ad on a particular publisher page is called an {\em impression}.  The registration of an impression by an advertiser's ad server is the starting point for payments that flow from advertisers through intermediaries to publishers.  The complexity of the ad serving process, the diversity of entities involved and the absolute amounts of money involved provide compelling motivation and opportunities for fraudsters.

The problem of fraud in online advertising has received increasing attention in both research and industry.  Typical fraud threats include but are not limited to {\em (i)} crawlers, traffic generators and bots that seek to inflate impression or click counts on publisher sites, {\em (ii)} browser extensions that make ad calls without a user's knowledge (ads are often shown, for example,  in pop under windows or invisible iframes) and {\em (iii)} ad injectors or other types of malware that insert unwanted ads on pages rendered by users.  Similar to other threats in the Internet, online ad fraud is a moving target.

In this paper, we describe a new ad fraud threat that we call {\em placement 
laundering}. We first publicly described this threat in the form a blog 
post~\cite{domainlaundering} where the attack was dubbed {\em domain 
laundering}.  Since that initial disclosure, we have identified other highly 
sophisticated instances of laundering. These discoveries have also suggested to us that 
the name {\em placement laundering} describes this threat more accurately.  We 
define placement laundering as intentional misrepresentation of the 
characteristics of an ad placement with the goal of drawing high priced ads to 
low quality placements that would otherwise draw only low priced ads from 
advertisers or exchanges.  The placement's quality refers to its prestige.
Consider that a high prestige placement might cost 
\$10 CPM (Cost Per Mille or per thousand ads) while a low prestige placement might cost \$0.01 CPM.

Placement laundering is facilitated by sending fake information along with ad 
requests to make low quality placements look like they are coming from higher 
quality publishers.  This is enabled by several different aspects of the ad 
delivery eco-system.  First, the ad delivery process makes an implicit 
assumption that the information that is sent from clients to ad serving 
infrastructure when a page is rendering is reported faithfully.  Second, the 
information available to advertisers is often limited by iframes and the 
diverse paths that an ad request can follow before it is finally served.  This 
takes advantage of cross-origin restrictions that all mainstream web browsers 
enforce.  Finally, there are no intrinsic capabilities in web or app 
infrastructure to assure or verify that an ad was delivered to a particular 
placement.  We call this the {\em attribution problem}.


Three key challenges stem from the attribution problem: detection, mitigation and prevention.
In this paper, our focus is on detection.  We provide technical background describing 
how attribution within the ad ecosystem works. We provide a definition for placement laundering and then discuss details of two case studies of this type of fraud. The purpose of these case studies is 
to illustrate concretely the diversity of methods that can be employed and also
to highlight fundamental weaknesses in today's attribution mechanisms.
We report that one of the instances impacted 200,000 display impressions per day. The other, which is significant because it
abused the industry-standard SafeFrame environment, served between 1M and 5M impressions per day.  
Neither malware nor special plugins are required to be installed on the client machine in order 
for either of these schemes to effect placement laundering. 

In Section~\ref{sec:detectmeasure} we describe details of a process and implementation that has proved capable of detecting multiple unrelated placement laundering schemes. To the best of our knowledge, there is no prior published record of a process that is capable of generalizing from an individual example of placement laundering.
We apply our detection method to a corpus of 434B 
publisher page view events and 31B ad impressions that were collected during March 2018.  We report that our process identifies five
distinct laundering schemes.  A critical feature of our implementation is that, although our process is informed by
proprietary data sources, an equivalent process could be built using standard data 
sources such as network-level packet traces and WHOIS queries.  
Validation of our implementation is informed by instrumentation 
deployed across 2M residential desktop machines. In addition to providing us with ground truth, 
this data allows us to review functional details about individual schemes. 
Using this data, we are able to distinguish several unrelated laundering schemes. We also provide new details  about an instance that was first reported in November 2017 called Hyphbot.

Regarding the other two key challenges, namely mitigation and prevention, we argue 
that these are likely to remain major challenges for fundamental reasons.
Addressing the attribution problem directly offers an opportunity to significantly diminish 
this threat.  One approach would be to require trusted entities to collaborate 
at each step of the ad serving process toward the goal of verifying that an ad 
was delivered to a specific publisher page and shown to a specific user.  While 
there would certainly be challenges in creating both the business relationships 
and technical capabilities, the benefits would be significant.  




%% file: background.tex
In this section, we provide an overview of the online ad ecosystem.   While 
similar descriptions have been provided in prior studies of ad fraud ({\em 
e.g.,} ~\cite{Springborn13,Dave12,Gross11,Zhang11}) we highlight details that 
are of particular importance in our description of placement laundering in 
Section~\ref{sec:pl}.

\vsubsection{Entities in the Ad Ecosystem}

At a high level, there are four major participants in the online ad ecosystem.

\begin{enumerate}
\item {\em Users} who view web pages or apps via desktop or mobile devices.
\item {\em Publishers} who create and maintain web sites or apps that include space reserved for ads.  
\item {\em Intermediaries} who provide a wide variety of services that assist in the delivery ads.
\item {\em Advertisers} who seek to market their products or services by running ads on web sites or within~apps.
\end{enumerate}

Users are the most familiar of these entities.  Their primary role in the ad 
ecosystem is to initiate requests and to view publisher content and ads in 
browsers or apps\footnote{For the remainder of this paper, we will only refer 
to {\em web pages} viewed in browsers, however all of the issues apply in 
fairly equal measure to ads served to mobile apps.}.  The user's browser also 
plays a central role in what are potentially a large number of HTTP 3xx 
redirects between the time that an ad is requested and finally served.

Publishers maintain web pages that have space allocated for content and ads.  
The term {\em content} is used to describe all manner of information consumed 
by visitors. Examples include text-based news, images, videos as well as 
widgets that are designed to engage users or drive revenue. Publishers 
typically adhere to established standards to display ads. In the common case, 
the space allocated on a publishers page is a standard size ({\em e.g.,} 728x90 
or 300x250, where units are in pixels and the size is standardized by the IAB). 
The ad {\em creatives} (text, display or video) are designed to be rendered 
within viewports of these sizes.  Another common feature is extensive use of 
iframes to host ad creatives. Publishers also use {\em ad servers} to decide 
which ad source should be used when an ad request is initiated.

Intermediaries are entities that drive revenue for publishers and other 
intermediaries by offering ad management services, information services, 
monitoring services, fraud detection services, delivery services, {\em etc.}  
The number and diversity of these entities is dizzying and ever-changing as is 
illustrated in the well known and continually evolving Lumascape 
diagrams~\cite{Luma}.  Critically, information flow between intermediaries 
during the process of serving an ad can create opportunities for fraud 
including placement laundering.

Advertisers are companies that are paying the bills for the entire ecosystem.  
Money flows from advertisers through intermediaries to publishers where every 
entity involved in the ad serving process takes a percentage.  Big advertisers 
work with their agency partners to define {\em campaigns} that include an 
online component.  Campaigns can target user demographic segments, geographic 
areas or behavioral profiles.  In practice, campaign targeting is tied closely 
to publishers:  higher prices will be paid to generate impressions on web sites 
that are considered high quality and prestigious (either in general or in 
specific interest segments).  The differential in prices paid to place ads on 
high quality versus low quality web sites can be several orders of magnitude 
and is at the core of placement laundering. 

\vsubsection{Delivering an Ad to a User}

\begin{figure}[ht]
\centering
\includegraphics[height=2.5in]{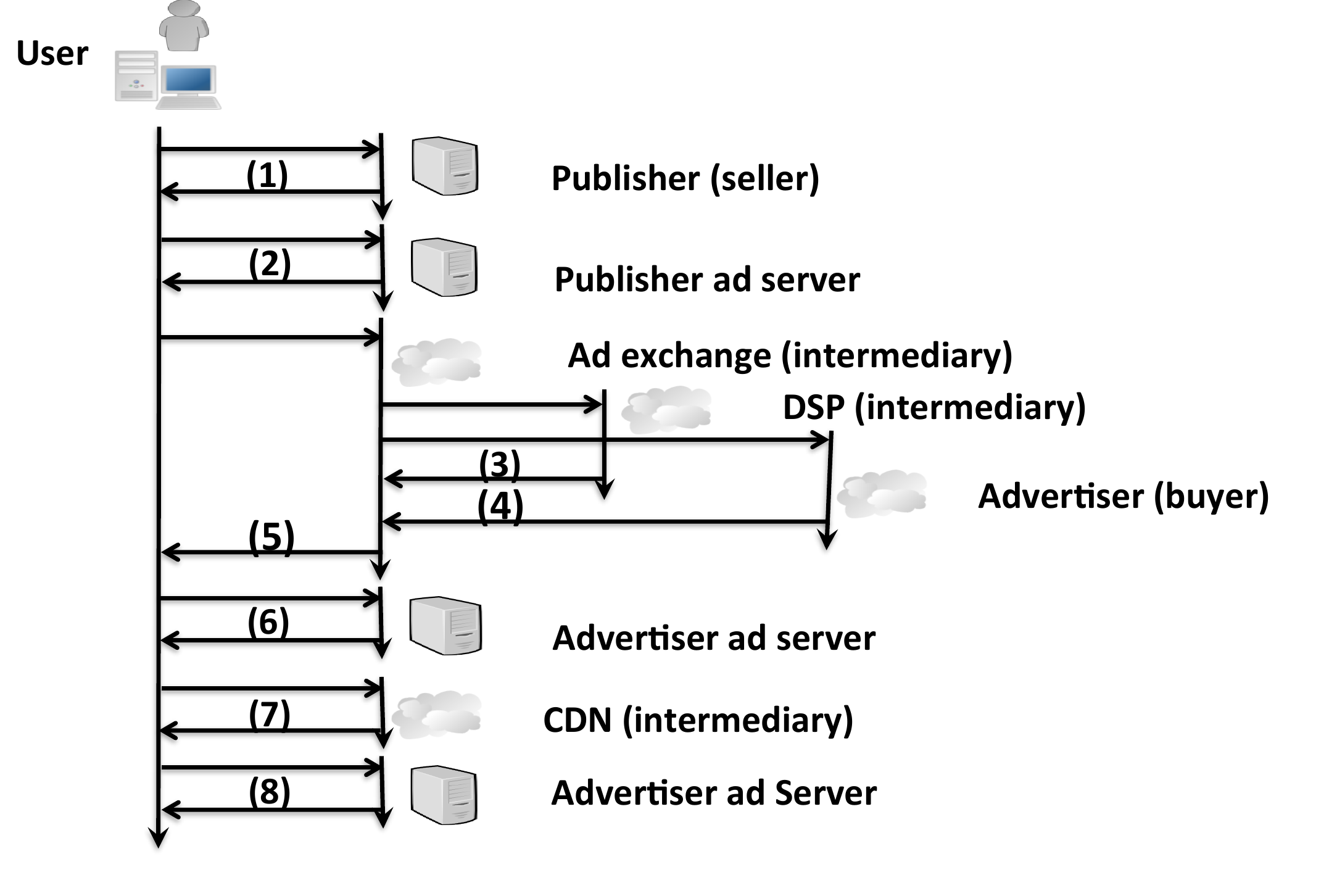}
\caption{\small Data flows and entities involved in delivering an ad creative in response to a request from a user.}
\label{fig:addataflow}
\end{figure}

Figure \ref{fig:addataflow} provides an example of data flows that result in an 
ad being served to a client.  In practice there are many different ways in 
which entities can participate in ad delivery and this example only illustrates 
one set of critical data flows.

The process begins with a User requesting a web page from a Publisher.  HTML is 
returned with embedded URLs including those for ad placements (1).  Ad requests 
are forwarded to the publisher's ad server (Publisher ad server), which will 
select among a number of potential options for delivering the ad.  When the 
choice is made, the server responds with a 302 redirect (2) to the selected ad 
source, which in this example is an ad exchange (Ad exchange intermediary).  
The ad exchange can make a number of different decisions about how to respond, 
including offering the placement to other entities such as Demand Side Partners 
(DSP) that act on behalf of advertisers who participate directly 
on the exchange (Advertiser).  These entities respond with bids to fill the 
placement (3,4).  The ad exchange makes the final selection and responds with a 
302 redirect to the selected entity (5), which in this example is the 
advertiser's ad server. The advertiser's ad server selects the appropriate ad 
creative and responds with a 302 redirect to the Content Delivery Network (CDN) that has the creative 
locally cached and a 302 redirect back to itself.  The CDN responds with the ad 
creative (7).  The second redirect back to the advertiser's ad server is used 
to register the fact that the ad was served which results in the placement of a 
tracking pixel on the client (8).  It is important to note that a page view 
is registered by the publisher when it serves the HTML in step 1, while an 
impression is only registered by the advertiser when the tracking pixel is 
placed in step~8.  These data are used to reconcile billing, which takes place 
after ads have been served.  The multiple redirects in this example also enable 
intermediaries to place cookies and thereby monitor transactions for billing 
and other purposes.

\vsubsection{Information Flow Challenges}
\label{sec:pl}

There are several aspects of the ad serving ecosystem that inhibit faithful 
communication between ad-serving entities.  First, due diligence by publishers 
and agencies who they partner with varies widely and in some cases is 
nonexistent.  Second, the ability to detect fraudulent views on web pages 
critically depends on the publisher's site structure as well as the publisher's 
motivations.  Third, there are inherent difficulties in information flow from 
client browsers to ad entities.  

\begin{figure}[ht]
\centering
\includegraphics[width=2.75in]{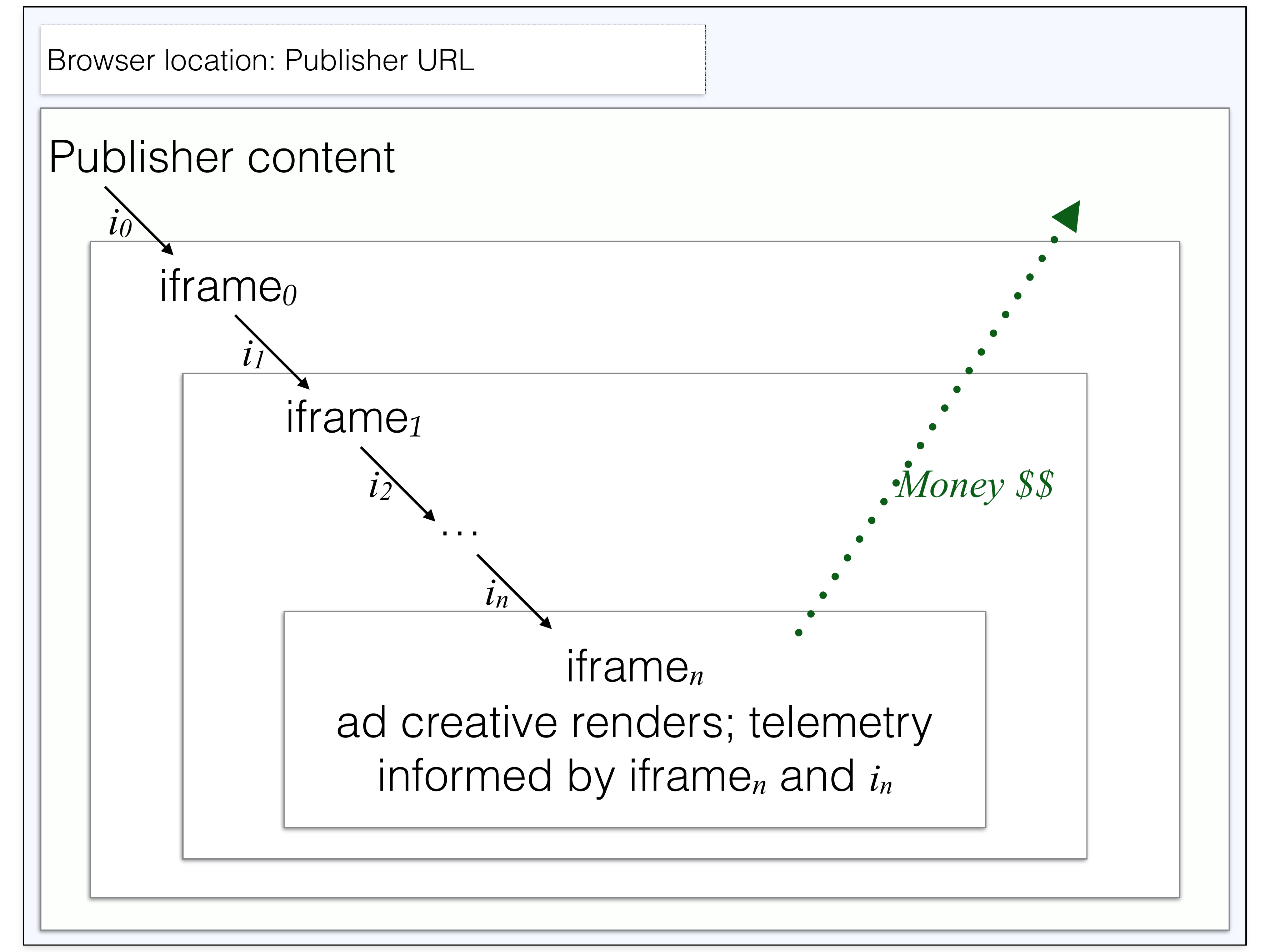}
\caption{\small The sequence $i_0\rightarrow i_1\rightarrow\cdots\rightarrow i_n$ 
abstractly describes the flow of information within the DOM during the course 
of delivering an impression.  Ad creatives render and campaign telemetry often 
executes within an iframe that is nested within other iframes.  Attribution to 
the publisher is possible only when the publisher's identity travels between 
$i_0$ and $i_n$ without alteration. Case study \ref{cs:safeframe} (SafeFrame) 
offers an illustration of how this is exploited. The flow of money is 
generally opposite the flow of information.}
\label{fig:nested}
\end{figure}

Consider Figure \ref{fig:nested}, which 
highlights how nested iframes limit information flow.  If code 
running alongside an advertisement within the innermost iframe attempts to identify the domain that served the outermost frame's 
content, the result is likely an error. 
Verification of the publisher is not 
generally possible due to browser cross-origin policies.  
Rather, the information that describes the ad's placement must be 
dutifully passed along, as represented by the chain of $i_j$'s, by all the intermediate parties. 
In fact, the only interaction that the Advertiser's ad server has is with the client -- not with the publisher. 

\begin{figure}[ht]
\centering
\includegraphics[width=3.0in]{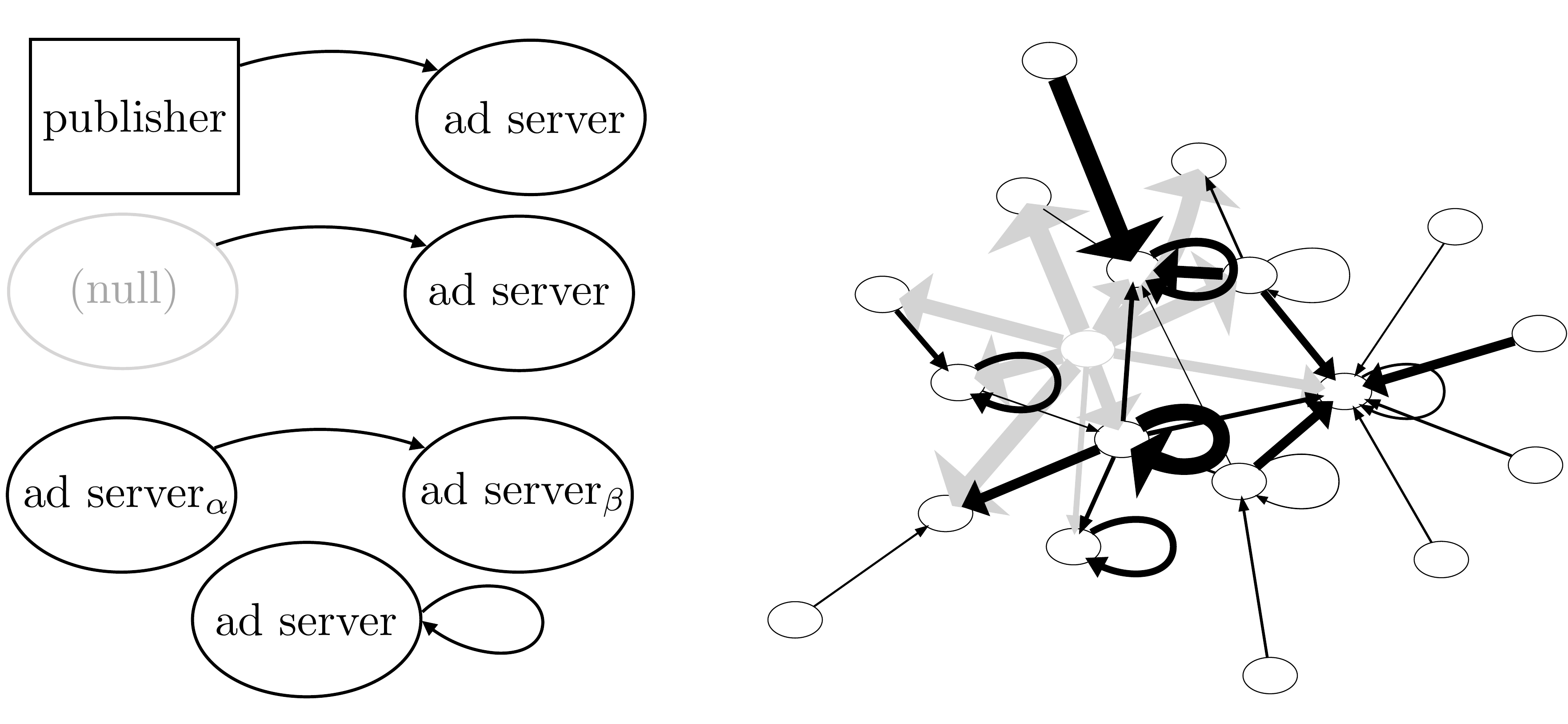}
\caption{\small An ad impression that is delivered by an ad server has one of several
classes of referer: empty, a publisher, or an ad server. 
At right is a graph of inter-adserver traffic based on one day of observations. 
This view summarizes 5B ad impressions. 
Each node corresponds to an ad serving platform (the names of
the servers are deliberately omitted).  Line width on this graph roughly corresponds to the volume 
of traffic.  On this day, about 1/3 of 
all traffic measured (about 1.5B impressions) reported a referrer that was an ad server.  
Loops represent traffic between accounts on the same platform.}
\label{fig:adservers}
\end{figure}

The relationships and information flow amongst ad serving entities 
are also complex.  Figure~\ref{fig:adservers} highlights interactions that 
commonly occur among large ad serving entities.  This figure shows that the volume of inter-ad server traffic is highly complex.
All the traffic represented in this figure is at least one step removed from the original publisher's domain.  
The view presented is based on instrumentation deployed on over 5B ads 
delivered by major ad serving entities over the course of a day.   

To summarize, accurate measurement of attribution is a challenge for 
fundamental reasons.  Attribution within the ad ecosystem 
currently relies on trusting the veracity of information that has been 
passed along by an unknown number of third parties.  

%% file: stdmeth.tex
In this section we describe attribution methods that are commonly employed in 
the ad ecosystem today.  We also note the limitations of these methods, which 
offer opportunities to fraudsters. 

\vsubsection{Standard sources of information and their veracity}

A common technique for collecting ad campaign telemetry is to send a snippet of Javascript code called a {\em tag} along with the ad creative when it is delivered to a client.  When the tag executes in the client browser, it inspects the local environment and requests the URL of a pixel appended with query parameters ({\em i.e.,} steps 7 and 8 in Figure~\ref{fig:addataflow}).  This pixel, called a {\em tracking pixel}, may be served from a host operated by a CDN and is rendered invisibly by the browser.  Campaign reporting and attribution is often {\em completely dependent} on the information in query parameters.

The Javascript code deployed for the purpose of gathering campaign data must be 
carefully engineered to run on a wide variety of browser and host 
configurations.  While Javascript is widely supported, it is also expected that 
the path of execution depends on the browser and host configurations.  Thus, 
tag development entails significant testing to ensure that the code will run as 
designed when deployed on a broad set of configurations. It has been shown that 
even the most elementary functionality can, despite careful engineering and 
testing, result in unexplained systemic anomalies~\cite{integratag}.

Javascript telemetry can provide a wide range of information about a client and 
context of an ad, {\em e.g.,}~details of user actions, the user environment, 
page properties and even portions of the raw HTML page.  Unfortunately, 
identifying the publisher page on which an ad is placed requires information 
from the top-level browser page.  This information is often inaccessible
as illustrated in Figure~\ref{fig:nested}. 

Another potential source of attribution information is the HTTP header.  
Raw header information is not always preserved or provided by industry participants 
and even when it is, the information it contains is not necessarily accurate.  
HTTPS traffic will not forward the 
referrer field to non-secure HTTP URLs.  Facebook uses ``link 
shims''~\cite{facebook} to allow attribution of traffic to Facebook while 
protecting users from revealing personal information to third parties.  A W3C 
proposal~\cite{w3crefpolicy} further refines what page authors can include in 
HTTP headers.  The RefControl~\cite{refcontrol} browser plugin is designed to spoof the \texttt{Referer} field.  
As demonstrated in the case study included in the 
appendix, even if the referrer and URL are accurately reported, they can be 
misleading.

Finally, in order that payments for a delivered impression flow and that everybody involved in the delivery process
receives credit, account identifiers and tag identifiers that are presented to ad exchanges, DSPs and other sources of ad inventory 
must be accurate.  This is true even when the traffic is otherwise misleading.

\vsubsection{Placement Laundering}
We define placement laundering as intentional misrepresentation of the 
characteristics of an ad placement with the goal of drawing high priced ads to 
low quality placements that would otherwise draw only low priced ads from 
advertisers or exchanges. To better understand how economic incentives entice fraudsters
to engage in domain laundering, consider the following.

Companies that buy ad campaigns often require that their advertisements avoid
placement on sites that host controversial or offensive content, {\em e.g.}, hate speech.
But if a fraudster is able to substitute an ad's correct attribution 
on such a site with the name of an innocuous-sounding domain, then laundering is occurring. 
With the true placement cloaked, the perpetrator is free to engage with a larger set of ad serving platforms and ad campaigns.
In fact, the scheme that is described in the Appendix placed ads on adult sites while the domain of attribution 
misleadingly suggested that the ads were delivered on sites about home repair and gardening.


\vsubsection{On Validating Placement Laundering Schemes}
Validating instances of laundering requires supplementing misleading telemetry with
ground truth.  As already discussed, the simplest idea, namely to deploy 
JavaScript alongside a brand's campaign creatives and to instruct the 
JavaScript to inspect, and then report about the top-most element of a page's 
DOM, does not work.  This is because advertisement creatives are generally 
rendered within iframe elements and cross-site scripting policies, which are fundamental to the browser security 
model, restrict what is observable from within such an environment.  

An alternative idea is to instrument both advertiser campaigns as well as 
publisher pages with JavaScript-based telemetry. Analysis of the two data 
streams reduces to reconciling the logs.  The publisher census and the ad 
campaign tags that are deployed by comScore appear to provide the proper 
perspective to do this. However, our experience is 
that this approach results in an unacceptably high false positive rate.  We conjecture that reasons
include browser handling of third-party 
cookies, the widespread use of ad blockers, non-comprehensive deployment of
telemetry by publishers and technical complications that arise from 
advertisements that load and refresh asynchronously with page views.

Since general automated detection is still an unsolved problem, we establish ground truth 
via a forensic, manual review of suspected instances of laundering. 
This can involve a direct 
manual review of all three data types (panel, census and ad campaign - explained below) as well 
as direct observation of browser behavior. This latter technique often implies an 
inspection of browser HTTP traces of suspect URLs.  Despite the obvious limitations 
of this technique, it is a rich source of information to either confirm or refute the 
presence of fraud. 

%% file: casestudies.tex
\label{sec:cs}
Placement laundering attacks can happen in many ways, ranging from simple 
referrer spoofing to far more complex methods where malware alters system 
parameters.  Of the many cases of placement laundering that we have verified, in this section
we describe two illustrative instances.  The examples 
are very different from each other, though the common feature is that in each 
case, an ad creative's attribution was falsified.  A third case study that is 
especially complex and involves misattribution in video ad delivery is 
described in the Appendix. Note that malicious software 
is {\em not} required to be installed on the client in order to execute 
any of these schemes.
\input{datasrc}
\input{adhexa}

\input{safeframe}

%% file: datasrc.tex
\vsubsection{Our data sources}
Three types of data inform our view of Internet traffic: publisher page view 
events, ad campaign impression events, and panelist web traffic records.  Each 
plays an important role in the process of detection and verification of 
placement laundering.  We describe each of these data types, why each is 
collected, and include high-level statistics to provide a sense of scale.

\paragraph{Publisher census}
Publishers often partner with third parties to measure audience demographics and reach and 
also to participate in publisher ranking reports.  Participation is straightforward and 
requires including a Javascript tag or the URL of a tracking pixel on their pages 
which fire when a browser renders the page. 

comScore's publisher census participants include 90 of the top 100 US-based publishers and thousands of 
other domains.  Census data typically records over 14B events daily. The information 
from this source that was used in our study includes timestamp, URL, referrer, and cookie.

\paragraph{Ad campaign tags}
Similar to publishers, companies that run ad campaigns partner with third parties 
to verify audience reach and other features of ad campaigns.  JavaScript tags are 
delivered with ads that appear in placements on publisher pages.  Impression 
delivery and campaign accounting must be done in accordance with 
well-established industry standards concerning viewability, brand 
safety and audience targeting~\cite{IAB}.  As a result the typical ad tag is a fairly 
complex JavaScript program. The information returned to the third party is 
often a rich description of the browser and context in which the code executed.
comScore deploys its own ad campaign telemetry, the collected data 
exceeds~1B impressions per day.


\paragraph{Panel data}
\label{sec:panelist}
User panels are another way in which third party measurement groups collect 
information on publisher and ad campaign activity.  comScore maintains a panel 
of~2M user participants~\footnote{comScore is highly sensitive about user privacy issues. Policy details can be found at http://www.comscore.com/ About- comScore/Privacy}. Each panelist provides detailed personal 
information and agrees to install monitoring software, called the Meter, on her host machine. This 
software can report all HTTP(S) traffic to/from their host machine. It also 
reports detailed diagnostic information such as process 
names and other identifiers to comScore.



%% file: adhexa.tex
\vsubsection{Adhexa}
Our first case study is a simple scheme that was discovered during May 2015.
URLs of ad calls posted to ad exchanges can contain information in a human readable format.  
Misattribution occurs when the query parameters that represent the publisher's domain are false.  

\begin{figure}[tp]
\begin{center}
\includegraphics[width=4.25in]{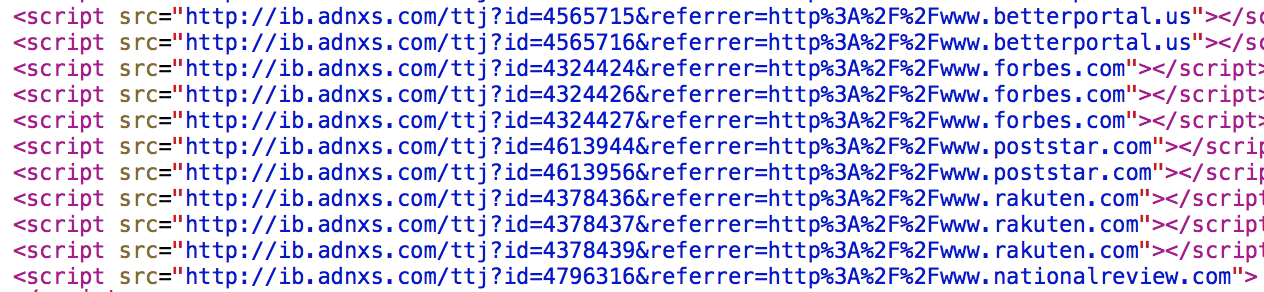}
\end{center}
\caption{\small This figure shows a screen capture of code that is delivered during the course of a single ad call on the Adhexa network. Related code instructs a web browser to create several sibling iframes, each of the siblings instructs the browser to request another ad. Note that the query parameter ``referrer'' in these lines cannot all possibly be accurate.} 
\label{fig:admediaco-science-forbes}
\end{figure}

The scheme begins when a browser requests an ad from an ad exchange and is returned a 
URL that displays a creative for the ad nework, {\em Adhexa}. The code 
displayed in Figure~\ref{fig:admediaco-science-forbes} also loads.  This 
activity occurs within an iframe that lives on a publisher's page.  Additional 
code (not shown) appends several hidden iframes to the page's structure.  
Ephemeral domains used to serve the hidden iframes include
\url{nxsrv1.com}, \url{admediaco.science}, and \url{miley-music.cf}.
These domains were only active for several weeks around the time of discovery.
Each hidden iframe initiates an auction on yet another ad exchange with a ``referrer'' query string parameter randomly selected from a predefined list. 
Several different account identifiers are used, so the risk born by any one account on a single platform is limited.


We estimate the flow of money associated with this single instance is several 
thousand to tens of thousands of dollars per day. This is based on a CPM in the 
range of \$0.01 to \$0.10 and the measured impression volume of 200k to 500k per 
day.

%
%

%% file: safeframe.tex
\vsubsection{SafeFrame}
\label{cs:safeframe}
The second case study concerns SafeFrame. SafeFrame is the
industry-standard API-enabled iframe whose functionality is 
defined by the Interactive Advertising Bureau (IAB).  Its purpose is to allow 
information to pass between ad creatives and publishers in a secure and 
standardized manner.  Google's own implementation of SafeFrame, which was the target of this scheme, is integrated 
with their Doubleclick for Publishers platform. It has extremely wide 
deployment and is described in the Google Publisher Tag library documentation.  
This environment is host to billions of advertisements per day.  The initial 
public disclosure of this issue occurred in close coordination with the Google 
Traffic Quality team\cite{safeframe}.



This attack was active during January 2016. It abuses the SafeFrame's
intended functionality to load and execute malicious JavaScript hosted on a 
third-party server.  When executed, this code repeatedly creates and destroys 
iframes within the SafeFrame. Each new iframe hosts an ad but it also has critical methods
belonging to the iframe's global {\tt window} object overwritten.  Specifically, each iframe is 
created with the URL-encoding methods \texttt{escape}, 
\texttt{encodeURI} and \texttt{encodeURIComponent} redefined. The redefined 
methods inspect whether the passed-in argument matches Google's SafeFrame URL.  If the 
argument matches, the returned value is selected from a list of URLs that are 
loaded dynamically during the course of ad delivery. If the argument does not 
match, the original encoding method is called and the standard behavior occurs. 
Figure \ref{fig:safeframe} illustrates this malicious functionality.

The implications of this attack are significant.  {\em Any tag that executes 
within the tainted environment and does not first verify the integrity of 
global JavaScript methods is at risk. } Based on direct measurments of campaign 
data, we observed daily volume varying between 1M and 5M impressions per day 
over a 30-day period.
\begin{figure*}[tp]
\begin{subfigure}{\textwidth}
\begin{Verbatim}
> escape + "";
  "function escape() \{ [native code] \}"
> escape("http://tpc.googlesyndication.com/safeframe/1-0-1/html/container.html");
  "http%3A//tpc.googlesyndication.com/safeframe/1-0-1/html/container.html"
\end{Verbatim}
\label{fig:cleanenvironment}
\end{subfigure}
\begin{subfigure}{\textwidth}
\begin{Verbatim}
> escape + "";
  "function(n) \{ return privateEncode(n, wrapper['escape']);\}"
> escape("http://tpc.googlesyndication.com/safeframe/1-0-1/html/container.html");
  "http%3A//www.example.com/high-quality-placement"
\end{Verbatim}
\end{subfigure}
\caption{\small The SafeFrame attack overwrites methods that belong to the global JavaScript {\tt window} object. Tampering with these methods can be detected by comparing the result of casting each method to a string. At top is JavaScript console output  showing the default global {\tt escape} method when cast as a string. It also shows the result of passing the Google SafeFrame URL to this function.  The bottom view shows similar commands executed within a tainted environment similar to what happens during the SafeFrame attack. The default escape function is redefined to invoke a function called {\tt privateEncode}. The default encode method is preserved as a fallback mechansim.  All JavaScript tags that invoke the standard functions {\tt escape}, {\tt encodeURI} or {\tt encodeURIComponent} without first checking their integrity are vulnerable.}
\label{fig:safeframe}
\end{figure*}


%% file: detectmeasure.tex
We now describe details and results of a process that is capable of 
detecting a general class of placement laundering schemes.
Many examples of 
laundering exhibit characteristics that are trivial to identify, but only after the 
instance has been discovered.  Designing a process that is able to identify new
schemes without relying on {\em ad-hoc} manual forensic investigation 
has proved to be a significant challenge.
To our knowledge, no prior published work describes a process that generalizes.

\vsubsection{Background}
Our methodology targets a general class of placement laundering instances.
A representative from this class was reported in November 2017, 
which the popular press dubbed ``Hyphbot'' \cite{hyphbot, adformhyphbot}. 
In terms of scale, press reports state that Hyphbot was facilitated by a fleet of 500K machines that 
were infected with malware and that it cost the industry \$500K/day.  
In Section~\ref{Sec:Results} we discuss our findings on Hyphbot and other instances that, to our 
knowledge have not been reported before.

The aim of our detection method is to identify ad attribution fraud occurring on 
non-mobile ({\em e.g.}, desktop, laptop, tablet) devices. The data used to inform this 
process is collected by the panel. Recall that the panel is facilitated by 
software that is voluntarily installed with informed consent by users on their 
computer. Once installed, the software collects and reports all HTTP(S) traffic 
that occurs on the client machine to comScore. 
The information stored with each HTTP(S) request includes the client-side outgoing 
request as well as the server response code, if any.  Additionally, the panel software
records the name of the process that originated the HTTP request and it records 
the dot-decimal formatted IP address of the server that is contacted as part of the 
outgoing request.  


While our detection method relies on proprietary data,
our method could equally well leverage data sources other than the panel.  For 
example, for unencrypted traffic, similar signals would be observed by sniffing 
packet-level data on a network.  Indeed, with packet data, one can recover the 
source/destination IP addresses as well as the URLs that are passed in HTTP 
headers. Once this data has been processed appropriately, the subsequent 
analysis of the data to detect laundering would be identical.

We now briefly describe key aspects of attack.  
When a web browser initiates an HTTP request, the message is routed by mapping
the domain name appearing in the URL to an IP address.  This mapping usually occurs 
via the global Domain Name System~(DNS), though operating systems commonly
maintain a local configuration file ({\em e.g.}, \texttt{/etc/hosts}) that also maps 
domain names to IP addresses. The mappings in the local file have precidence over whatever the 
global DNS reports.  An attack on either the local configuration file or the global DNS that results in
high-value domains mapping to non-standard IP addresses will result in the same signature that we seek.

The theory of why our detection method works is that major publishers who 
support large audiences host their content on extremely robust infrastructure. 
This is time consuming to configure and expensive to 
operate. Thus, for both business and technical reasons, high-reputation 
publishers typically host their content on the hardware operated by 
a small number (often just one or two) Internet
Service Providers~(ISP). The publisher-ISP relationship changes very slowly over 
time.  Conversely, malicious actors can rent low-cost, easy-to-configure web servers using cloud-based services. 
Our methodology does risk flagging arbitrary traffic that is delivered from a CDN that uses IP addresses 
that are owned by multiple ISPs to serve content.  Our experience is that operationally, we are 
robust against this scenario.


\vsubsection{Method and Results}
\label{Sec:Results}

The essential idea of our detection method is to identify IP addresses that are associated with
multiple unrelated high reputation publishers by panelist machines.  This can happen for a variety of reasons including but not limited to a customized {\tt /etc/hosts} file or some other DNS hijack attack.
The output of our process is a key-value pair, where the key is the tuple (IP, ISP) 
and the value is a set of distinct domains that 
resolved as the IP address in the key by panelist machines.
We also store the name of the processes that initiated each HTTP(S) request.
We use process name as part of followup analysis to confirm or refute the existence 
of malware on an indiviudual machine. Observe that
the (IP, ISP) pair is the server-side of the HTTP request while the process name
is purely client-side ({\em i.e.}, it is not transmitted over the network).

Our method, which has been calibrated to achieve a low false-positive rate, runs as follows:
\begin{itemize}
\item A set of candidate domains is built.
    \begin{itemize}
    \item Each domain is high value (this is defined below) and
    \item Each domain is observed resolving to several IP addresses and at least two distinct ISPs are represented. 
\end{itemize}
\item If an ISP owns an IP address that resolves to at least twenty high value domains that live
in the candidate domain set, then the (IP, ISP) tuple and
the set of high value domains that resolve to the IP are among the key-value pairs of the output.
\end{itemize}
We label a domain {\em high-value} by referencing 
a list of publisher domains maintained by comScore. 
The Alexa ranking is similar and would likely generate similar 
results. We restrict our view to the top-ranked 2,000 domains. 
We find that around 325 high-value domains are typically affected at any point in time.

To label an (IP, ISP) key a {true detection}, we require that traffic associated with it
be generated by process names that are associated with known malware.  
This additional restriction means we identify only a subset of schemes, but we are highly 
sensitive to false positives and this step limits alerts to those that are likely to be true positives
We note that if a process has the ability to
alter protected domain resolution files ({\em e.g.}, {\tt /etc/hosts}),
 then it certainly can launch a separate process that has an inconspicuous name. 
\begin{table}[ht]
  \centering
    \begin{tabular}{lcccccc}
\toprule

      Label                        & \# ISPs & \# IPs & \Centerstack{\# Days seen,in March} & \Centerstack{\# of top 2K,domains} & \# Machine IDs & \Centerstack{Average \# of,daily requests} \\ 
\midrule
        Hyphbot                   & 4    & 25  & 31            & 726            & 141                   & 326,239                  \\ 
        $\mathrm{Scheme}_\beta  $ & 1    & 4   & 31            & 208            & 15,901                & 60,328                   \\ 
        $\mathrm{Scheme}_\gamma $ & 1    & 4   & 31            & 163            & 13,619                & 18,878                   \\ 
        $\mathrm{Scheme}_\omega $ & 1    & 1   & 30            & 364            & 134                   & 6,704                    \\ 
        $\mathrm{Scheme}_\lambda$ & 1    & 6   & 6             & 168            & 80                    & 1,611                    \\ 
\bottomrule
    \end{tabular}%
  \caption{\small We identify five schemes by applying our process to the March 2018 data snapshot. High-level statistics about each scheme are shown.}
  \label{tab:launder}
\end{table}

We apply our methodology to panel data that was collected across all 31 days of March 2018.
We also undertook a manual review of HTTP(S) requests originating from a small number 
of individual panelist machines.  The purpose of this review is exploratory and aims to identify the 
signatures of laundering schemes. 

Over the course of a day, our methodology typically flags $O(10)$ candidate (IP,~ISP) pairs.
The set of pairs detected by our process is fairly stable for timescales of about~1 week. 
We observe the ISP space is more stable than the IP address space, which 
aligns with our understanding of how cloud services allocate their resources to their customers.
The set of ISPs identified by our process changes slowly, with $O(1)$ emerging or disappearing each week.
Manual review of the ISPs that are identified suggests that cloud-based 
infrastructure is being used. 

Instances of laundering are distinguished from each other by several features, including server-side
infrastructure, client-side process names, the User Agent field of the HTTP request and URL structure. 
Identifying the distinguishing features of laundering schemes is currently 
a manual effort. Automation of this would represent an enhancement, but is separate from our main goal of detection so we leave it to future work.
The purpose in manually distinguishing unrelated schemes here is to show that our detection method generalizes.
Within our data snapshot, we identify
five distinct instances of placement laundering. Table~\ref{tab:launder} displays a high-level summary.

\begin{table}[ht]
  \centering
    \begin{tabular}{lccccc}
\toprule
                  & Hyphbot & $\mathrm{Scheme}_\beta$ & $\mathrm{Scheme}_\gamma$ & $\mathrm{Scheme}_\omega$ & $\mathrm{Scheme}_\lambda$\\
\midrule
        Hyphbot      & 1 &  0.09 & 0.08 & 0.30 & 0.18\\ 
        $\mathrm{Scheme}_\beta$      & - &  1    & 0.32 & 0.07 & 0.04\\              
        $\mathrm{Scheme}_\gamma$     & - &  -    &  1   & 0.05 & 0.03\\
        $\mathrm{Scheme}_\omega$     & - &  -    &  -   & 1    & 0.39\\
        $\mathrm{Scheme}_\lambda$    & - &  -    &  -   & -    &   1 \\
\bottomrule 
    \end{tabular}
\caption{\small The Jaccard index between the sets of domains that are associated with each scheme is shown. 
Several schemes have nearly disjoint sets while others exhibit significant overlap. Overlapping schemes
are classified as distinct from each other for having other distinguishing traits such as process name, URL structure or the ISP used.
We only fill in the upper triangular portion of the matrix since, for any sets $A$ and $B$, the Jaccard index 
$J(A,B)=J(B,A)$, so the complete matrix is symmetric.}
\label{tab:jaccard}
\end{table}

To quantify the differences among schemes identified, we use the Jaccard index 
of the set of domains that are among the top 2K domains.
The Jaccard index is a statistic that applies to set pairs.
For arbitrary sets $A$ and $B$, the Jaccard similarity index between $A$ and $B$ 
is $J(A,B)\coloneqq \abs{A\cap B}/\abs{A\cup B}$.
The top 2K domains are all likely targets of schemes, so this metric alone is not 
sufficient to distinguish unrelated schemes from each other. 
That said, it is illustrative. Table~\ref{tab:jaccard} lists the Jaccard index between
the schemes we classified.

Schemes are often distinguished from each other via the executable process names that generate their traffic.
The executable(s) associated with Hyphbot and $\mathrm{Scheme}_\omega$ were recorded as either whitespace or 
empty strings.  We hypothesize that this is deliberate and that the process name 
is configured to evade detection.

We now compare our observations about Hyphbot to earlier public reports.
The regular expression patterns derived from URL lists that were described in the press~\cite{adformhyphbot}
match 1.5\% of the 
traffic that our method identified. We have observed schemes evolve over time, and
detection methods that are informed by URL lists or regular expressions are easily circumvented.
We are able to estimate 
impact as 250K impressions per day.
We arrived at this figure by matching URLs that resolve to Hyphbot infrastructure to 
direct measurements of campaign traffic.

We close this section by providing additional detail to highlight the unusual functionality 
of Hyphbot and $\mathrm{Scheme}_\beta$. 
Based on manual review of panel data, we believe that client-side tampering of the hosts file has occurred and that
misleading ad attribution is the result of this tampering.
Within $\mathrm{Scheme}_{\beta}$ traffic, we observe panelist machines requesting pages from malformed domain names
({\em e.g.}, {\tt  li.zulilycom}~[sic]) and we observe that name resolution of these malformed domains succeeds.
We conjecture that such names are the output of a misconfigured script that generated a list of
target domains, and this list was delivered to infected machines via a command-and-control server.
Regarding Hyphbot traffic, during the course of ad delivery we observe URLs with the query fields
{\tt spoof\_domain} and {\tt land\_ip}. We observe that when these URLs occur in traces of panelist 
activity, later in the trace we indeed observe requests to the domain in the {\tt spoof\_domain} query field and the 
panelist machine resolves the IP address listed in {\tt land\_ip}.

%% file: discuss.tex
A natural question is whether laundering detection can be accomplished in a more straightforward way. 
For example, is it possible to simply reconcile the records from a publisher's web log against ads that are attributed to that publisher? 
We have attempted a similar idea using data sourced from instrumentation deployed across
publisher sites.  But this theoretically sound approach has proved ineffective in practice 
due to ad blocker use, incomplete deployment of instrumentation by publishers and campaigns and
business rules (rather than technical rules) that inform attribution.

The detection method of Section~\ref{sec:detectmeasure} 
serves as a starting point for understanding the scope and characteristics of the placement laundering threat.  
In an operational setting, identification of any kind 
of fraud must be done in a conservative fashion since false positives
lead to inappropriate deletion of revenue for publishers and intermediaries.  
 Our automated methods are often coupled with manual 
forensic investigations to confirm specific instances of placement laundering.  
We note that our process is not unlike manual processes that are used to 
decompose malware and generate signatures for intrusion detection and 
prevention systems. 

While detection coupled with reconciliation is a process that might be made to 
work, we argue that addressing the core attribution problem directly would 
significantly improve the situation.  One way to do this would be to create a consortium of trusted entities with 
the common goal of verifying that an ad was delivered to a specific publisher 
page and shown to a specific user.  This would have to be 
implemented in a way that preserves efficiency in the ad delivery process and 
in a way that would make the last step in the process, namely verifying delivery to 
a specific user visiting a page, difficult to circumvent.  This may well require new 
capabilities in browsers and in systems used by ad serving intermediaries.  We 
believe that this or other approaches to addressing the attribution problem 
offer attractive opportunities for future work. We note that the recently
standardized {Ads.txt}\cite{adstxtmain} specification
is reliant on accurate hostname resolution, among other things. Participants in this program 
are not guaranteed protection against placement laundering or other threats\cite{adstxtslides}.

%% file: relwork.tex
Fraud in online advertising has been addressed in a large body of prior work 
over the past decade.  Click fraud received significant attention as 
search-based advertising grew in popularity.  These pay-per-click-based 
advertising systems led to a variety of threats including botnets that had 
specific capabilities for click fraud~\cite{Alrwais12,Miller11,Daswani07}.  
Examples of botnets with click fraud capability include the Bamital 
botnet~\cite{Kirk13} and the ZeroAccess botnet~\cite{Infosec13}.  Prior work 
has also focused on developing methods for identifying click-fraud {\em 
e.g.,}~\cite{Zhang08,Metwally05}.   Haddadi develops the idea of using {\em 
bluff ads} for measuring fraudulent clicks and to generate IP blacklists to 
mitigate the threat~\cite{Haddadi10}.  Similarly, Dave {\em et al.} develop 
novel methods for measuring and detecting click fraud in ad 
networks~\cite{Dave12,Dave13}. 

Other studies have used novel measurement methods or large data from online 
advertising entities to study fraud and identify new threats.  Zhang {\em et 
al.} purchase impressions for a honeypot website that they deployed and report 
on the characteristics of the resulting traffic~\cite{Zhang11}. Similarly, 
Springborn and Barford used a honeypot website to identify pay-per-view 
networks, which use pop-under windows to monetize 
impressions~\cite{Springborn13}.  Stone-Gross {\em et al.} report on different 
types of fraudulent behavior including impression spam based on analysis of 
logs from a large ad exchange~\cite{Gross11}. Thomas {\em et al.} conduct a 
large-scale empirical study of the impact of {\em ad injectors}, which are 
browser extensions that overwrite ads that would otherwise be delivered to 
users~\cite{Thomas15}.  Our efforts are similar to these in that we use logs of 
billions of ad impressions as the basis for our investigations.  To the best of 
our knowledge there are no prior research studies on placement laundering.

Threats against domain name resolution ({\em e.g.}, DNS) have been known for years as
documented by Atkins and Austein in~\cite{Atkins04threatanalysis},
Jackson {\em et al}.~\cite{jackson2009protecting}, Bernstein~\cite{bernstein} and 
Schneier~\cite{schneierLessons}. 
Since the HTTP protocol relies on the {\tt Host} header field to route messages,
HTTP traffic is reliant on DNS and similar services.  
The HTTP specification~\cite{rfc7230} acknowledges that attacks on
domain name resolution propagate upwards to threats against HTTP traffic.

Finally, as has been mentioned, ad fraud has received attention by the 
IAB~\cite{IAB_IVT} and in many reports in the popular press over the past 
several years.  Companies such as Google, Microsoft, Yahoo and others provide a 
variety of information about their invalid traffic monitoring 
activities~\cite{google_quality,bingads,yahoo}.  The motivation for this 
documentation is to reassure advertisers that players in the digital ecosystem 
are paying attention to threats.  Unfortunately, many intermediaries have no 
such documentation or have only recently begun to pay attention.  

%% file: sumry.tex
The huge amounts of money that flow though the online advertising ecosystem make it a compelling target for fraudsters.  The scale, diversity and complexity of the systems and processes that are employed to deliver online ads offer a variety of opportunities for fraud.  While some methods for fraud are well known and can be defeated through careful monitoring and detection techniques, new threats are always emerging.

In this paper, we report on a new form of ad fraud that we call placement laundering.  The objective of placement laundering is to offer ad placements on low quality sites that appear as if they are coming from higher quality sites in order to extract higher payments for each ad that is delivered.  If done effectively, this can result in an increase in payments per ad by several orders of magnitude.  The result is an increase in revenue without an increase in traffic. The basis for placement laundering is the fact that there are no inherent mechanisms for ensuring that an ad was delivered to its intended placement on a publisher page.  We call this the attribution problem.

We illustrate the details of placement laundering methods in two case studies. 
 These studies highlight how misrepresentations of placements can result in ad 
requests being attributed to higher quality publishers.  Among these case 
studies is one that exploits the functionality of the industry-standard API 
called ``SafeFrame''. Our case studies provide insights on the sophistication 
of the fraudsters that are utilizing placement laundering.

We describe and report on a placement laundering detection method.
We show this method is capable of identifying traffic from a general class of laundering schemes.
The method provides invaluable insight into 
the evolution and birth of different placement laundering schemes.
Importantly, it generalizes.
To validate our method, we report on observable characteristics particular to each scheme.
Finally, we report on the overall impact of the general class of placement laundering we target.

We conclude with a discussion of how opportunity for placement laundering might be diminished by creating a consortium of entities in the online ad ecosystem that collaborate to verify and attribute ad placements.  There are both business and technical challenges in this approach but the benefits would be significant.

In future work, we intend to work on the technical aspects of the placement attribution problem and to develop new detection methods.  We also continue to monitor the placement laundering phenomenon for evidence of new and emerging threats. iframe-based crawling offers an opportunity to investigate placement laundering in a generalized fashion without the need for specialized data.  We plan to investigate automated methods for identifying placement laundering via crawling in future work.

%% file: appendix.tex
This instance of placement laundering employs the most complex method of ad delivery in our survey.  The techniques used include ephemeral URLs,  strong encryption, heavily obfuscated javascript code and metaprogramming to execute video ad delivery.  Additionally this example exhibits scalability, durability and adaptability over time. 

\begin{figure}[ht]
\centering
\includegraphics[width=4.0in]{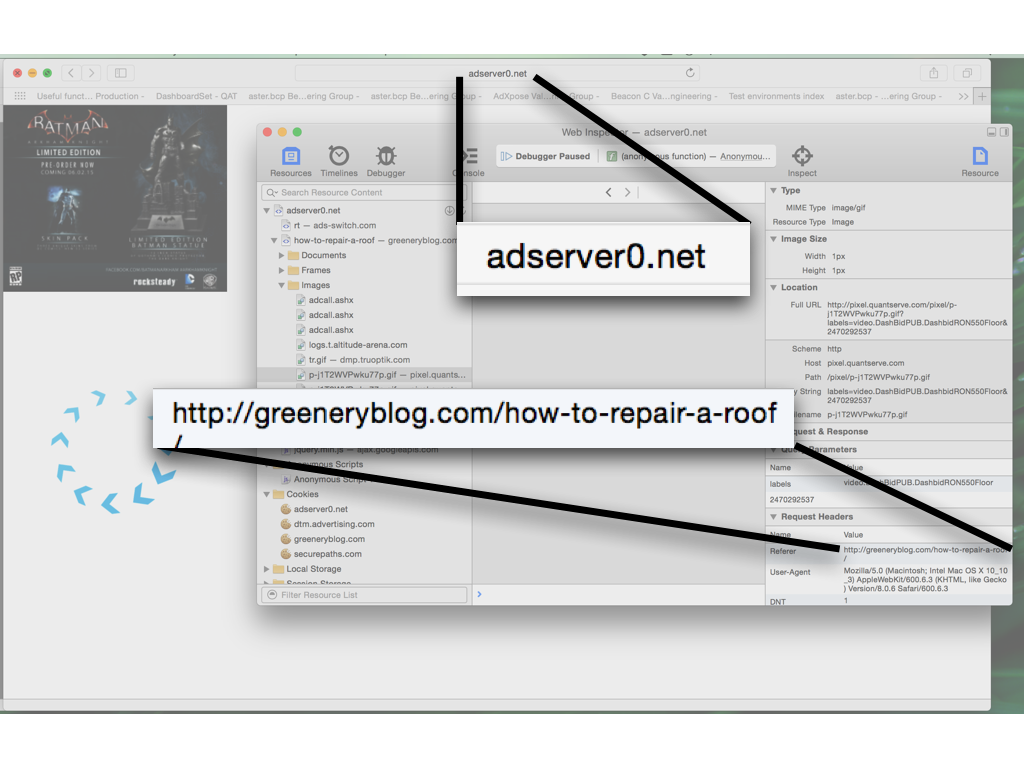}
\caption{\small 
	This screen capture, related to the bimage scheme, shows a video loading underneath an ad hosted on \url{adserver0.net}. The HTTP Referer field of the innermost iframe is set to a URL that concerns home repair projects. The domain \url{greeneryblog.com} and the story concerning home repair were not loaded by the user during the sequence of events that resulted in this impression delivery. What is visible here would normally render within an iframe on a third party publisher's page.}
\label{fig:ads-stream-net-human-readable-referer-2}
\end{figure}

The URL that ultimately returns a video ad to a browser is human-readable and suggests innocuous content. Several example domains are {\tt\url{loansaver.com}} and {\tt\url{trainsandplaneshobbies.com}}. Telemetry that accompanies a video ad that happened to be delivered in this manner will indicate that the ad was displayed alongside human-readable content on a family-friendly publisher's page.  See Figure \ref{fig:ads-stream-net-human-readable-referer-2}.  However, in early 2016, we associated between 2\% and 5\% of ads served on these domains with ad networks that cater to adult content. A followup review of panelist traffic during November of 2016 confirmed this. This is an issue called {\em brand safety}.

During late 2015, this scheme was associated with hundreds
of distinct domains.  We observed about five domains in service any point in time. We also observed that the set of highly-active domains changes regularly.  Daily impression volume varies, ranging from 200k video impressions on active domains per day 
 to over 1M video impressions per day.
 There are also dozens of ad servers whose purpose appears solely dedicated to the delivery of ads on these domains. Given the intricate, complex flow of traffic involved in each ad delivery, we believe that the ad servers and these domains are simply different parts of the same operation.  
The nickname {\em bimage} is derived from the URL path of the iframes that contain these ads~\footnote{In early 2014, a news story (http://adage.com/article/digital/ad-fraud-operation-fools-detection-companies-nets-millions/293929/) associated several domains related to those described here with ad fraud.}.

Most of these domains are simply not accessible from search engine results. This is highly unusual for a publisher who wishes to earn ad revenue from organic human traffic.  Yet the following lines of Javascript code are common to many of the identified domains. The code redirects a web browser to an error page if the page referrer does 
not contain the string ``\mbox{monkeysee}'':
\begin{verbatim}
if (g.indexOf("monkeysee")==-1 && f != g){
 return c.RefererSecurityBreached=!0, 
  d=!0, 
  window.location = 
   "http://{0}/referror?{1}".format(f, b), 
  !0
}
\end{verbatim}
The consequence is that if a user attempts to reach one of these domains from a set of search engine results, the page renders an error.

Another unusual feature is that many of these domains display fake ads.  This is not obvious from a cursory visual inspection.  Indeed, the creatives are sometimes from well-known brands and recent ad campaigns.  Clicks on these ad creatives send a user to an appropriate landing page ({\em e.g.}~the landing page for a Tide creative is Tide.com). But no tracking occurs.  The links are simply direct HTML links to brand home pages.  It seems clear that this is done to make the target sites appear to be legitimate.  Furthermore, the content on each site is related to whatever the domain name suggests. For example, {\tt\url{carsluxurious.com}} hosts stories and photos of fancy cars. 




The sequence of messages passed between client and server are likewise complex. The conversation has two significant features: it updates the user's browser by placing cookies on it and it updates the web servers by creating URLs that expire after a few minutes.
Without the proper cookie, an ad will not render. As a result, even if a person is able see a complete sequence of URLs associated with an ad call, that person will not be able to reconstruct the steps required to witness a video ad. Instead, the user will enconter a 404-not found page, a non-monetized ad in a bimage.aspx page or the publisher's ``content'' site.  

Finally, the ad's video player code is obfuscated in an unusual way.  The video player code returns a function which, in order to run, accepts arguments that are derived from the web browser's cookie and the page URL.  The function appears to generate a string of javascript which, after several rounds of deobfuscation, is executed with a call to \texttt{eval()}.  This call, in turn, creates several anonymous Javascript functions which load a video player in the browser. The starting and ending portions of the Javascript video player code are:
\begin{small}
\begin{verbatim}
var J5L=(function f(o,s){var j='',
  E=unescape('%13%1E*L*%246%01%08f%5...
       <skip 150k bytes> 
    ...FI%11%02"), "liOu...qrFIfrOjgT")());
\end{verbatim}
\end{small}
This video player is hosted at a short-lived URL. When the ephemeral URL expires, a lightweight dummy player is returned instead.
The ``real'' URL returns a player with a code-base that is approximately twice the size (in bytes) of the dummy player. 

